# Helical polariton lasing from topological valleys in an organic crystalline microcavity

Teng Long,[1] Xuekai Ma,[2] Jiahuan Ren,[1,3] Feng Li,[4] Qing Liao,[1,]* Stefan Schumacher,[2,5] Guillaume Malpuech,[6] Dmitry Solnyshkov,[6,7] Hongbing Fu[1]

[1]Beijing Key Laboratory for Optical Materials and Photonic Devices, Department of Chemistry, Capital Normal University, Beijing 100048, People's Republic of China
E-mail: liaoqing@cnu.edu.cn

[2]Department of Physics and Center for Optoelectronics and Photonics Paderborn (CeOPP), Universität Paderborn, Warburger Strasse 100, 33098 Paderborn, Germany

[3]Tianjin Key Laboratory of Molecular Optoelectronic Science, School of Chemical Engineering and Technology, Collaborative Innovation Center of Chemical Science and Engineering (Tianjin), Tianjin University, Tianjin 300072, P. R. China

[4]Key Laboratory for Physical Electronics and Devices of the Ministry of Education & Shaanxi Key Lab of Information Photonic Technique, School of Electronic and Information Engineering, Xi'an Jiaotong University, 710049 Xi'an, China.

[5]Wyant College of Optical Sciences, University of Arizona, Tucson, Arizona 85721, United States

[6]Institut Pascal, PHOTON-N2, Université Clermont Auvergne, CNRS, SIGMA Clermont, F-63000 Clermont-Ferrand, France

[7]Institut Universitaire de France (IUF), 75231 Paris, France

**Abstract**

Topological photonics provides an important platform for the development of photonic devices with robust disorder-immune light transport and controllable helicity. Mixing photons with excitons (or polaritons) gives rise to nontrivial polaritonic bands with chiral modes, allowing the manipulation of helical lasers in strongly coupled light-matter systems. In this work, we demonstrate helical polariton lasing from topological valleys of an organic anisotropic microcrystalline cavity based on tailored local nontrivial band geometry. This polariton laser emits light of different helicity along different angular directions. The significantly enhanced chiral characteristics are achieved by the nonlinear relaxation process. Helical topological polariton lasers may provide a perfect platform for the exploration of novel topological phenomena that involve light-matter interaction and the development of polariton-based spintronic devices.

**Introduction**

The concept of topology and the mathematics of topological invariants have been introduced into physical systems and provide a robust possibility to topologically protect the transport of electrons or photons along the edge of a structure against perturbations such as defects and disorder.[1,2] It has been demonstrated in a range of physical systems, such as two-dimensional electron gases,[3,4] cold atoms,[5] mechanics,[6] microwaves,[7] and optical systems.[8,9] Topology in photonics provides a chiral one-way channel for the transport of light. Topological photonics has been widely studied, ranging from optical waveguides,[8-11] resonances,[12] generation and propagation of single photons,[13-15] to topological lasers.[16-18] Recently, this research has also strongly focused on the study of topological photonics with non-equilibrium exciton-polaritons (simply called polaritons hereafter) in planar semiconductor microcavities,[19-21] where intriguing and distinctive topological polaritonic states have been reported. Polaritons are hybrid bosons, originating from the strong coupling of excitons and cavity photons.[22-24] The excitonic component of this symbiotic part-light part-matter particle allows strong interparticle interactions with resulting strong optical nonlinearities, which are notoriously hard to achieve in pure photonic systems.[24] This kind of interaction provides additional degrees to steer polaritonic bosons and may open new avenues towards the observation of novel topological phenomena in non-equilibrium polaritons. In fact, the realization of topological polaritonic states has been proposed through the independent manipulation of the excitonic or photonic component,[25] such as the excitons in a topologically non-trivial system,[26] the photonic spin-orbit coupling (SOC),[18] and the photon-exciton hybridization process.[19]

Lasers, especially circularly polarized lasers, have attracted great attention in nanophotonics,[27] quantum optics,[28,29] and biophysics.[13,30,31] The reports on organic circularly polarized lasers mainly focus on organic chiral molecules[32] or helical microstructures.[33] However, there are some difficulties in the time-consuming molecular synthesis and only single-helicity laser emission was observed in a left-handed or right-handed helical materials. Notably, topological wave guides naturally support helicity-dependent light transport along separate one-way channels.[11] Thus, topological states offer two advantages for lasers: the robust topological protection from defects and disorder in the cavities, where the laser develops, and the controllable chirality of laser emission. Recently, a chiral propagation of the polariton condensate was experimentally achieved in a two-dimensional honeycomb lattice with nontrivial topological bandstructure, originating from the photonic SOC and the Zeeman splitting of excitons under strong magnetic fields, at 4 K temperature.[18] However, the operation of strong magnetic fields, the low temperature requirement, and difficulty of the lattice fabrication hinder the practical application of such topological devices. Fortunately, it is possible to avoid these challenges in a photonic valley Hall system, such as in an organic polaritonic microcavity.[34] The larger exciton binding energy of organic Frenkel excitons facilitates the strong coupling of excitons and cavity photons at ambient conditions. Moreover, organic anisotropic crystals allow local nontrivial band geometry for an optical valley Hall effect without the design of specific lattices.[35] However, so far, controllable chiral polaritonic lasers based on these topological valleys remain unexplored.

Here, we demonstrate helical polariton lasing from topological valleys in an organic anisotropic microcrystalline cavity. Experimental and theoretical results indicate that light of opposite helicity is emitted along different angular directions. In this case, strong interaction of the excitonic component ensures polariton relaxation and lasing, while the photonic component offers topologically nontrivial valleys. The left-handed and right-handed circularly polarized lasing is topologically protected by the valley Chern numbers, which originate from the optical activity of crystal structures. The high helicity of output light is achieved by the nonlinear condensation process. Our findings in this work provide key insights into helical topological polariton lasers and may lead to applications of topological organic laser devices in the strong coupling regime.

**Result and discussion**

**Selective strong coupling in an organic microcavity.** Our organic microcavity consists of an organic active layer sandwiched between two metallic films as sketched in Fig. 1a. The active layer consists of single-crystalline microbelts of an organic molecule, 1,4-dimethoxy-2,5-di(2,2',5',2''-terthiophenestyryl)benzene (TTPSB). The details on the synthesis of TTPSB molecules, the preparation of microbelts, the fabrication of microcavities and their characterization are provided in the Supplementary (see Materials and methods). In our case, the chosen microbelts have smooth surface and uniform morphology with typical width of around 50 μm, thickness of 720-730 μm and length of around several hundreds of micrometers (Supplementary Fig. S1). These samples exhibit strong anisotropy along the axes of the microbelt crystal (i.e. X-direction and Y-direction) as shown in Fig. 1a. We first analyze the unpolarized total reflectivity of the sample in X-direction as shown in Fig. 1b, where the reflectivity is plotted as a function of wavelength (or energy) and angle (or momentum $k_x$), measured by using a home-made angle-resolved microscope at room temperature (see details in Supplementary).

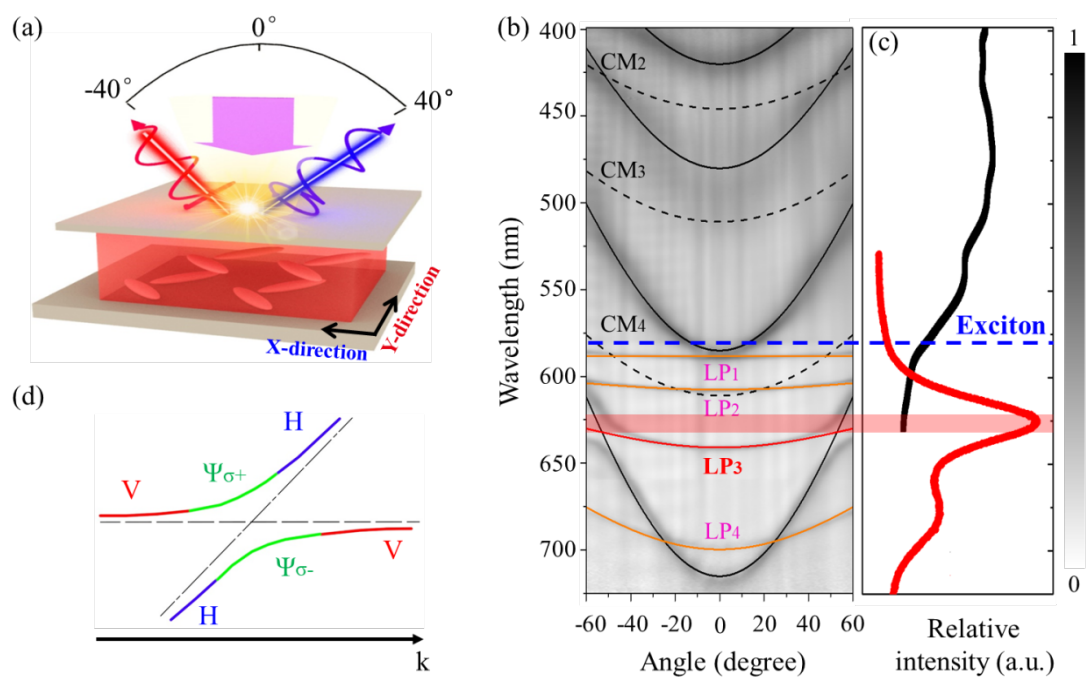

**Figure 1. Schematics of organic microcavity setup and angle-resolved reflectivity.** (a) Scheme of a TTPSB microbelt-filled optical cavity sandwiched between two silver reflectors. (b) Angle-resolved reflectivity of the microcavity at room temperature. $LP_n$ and $CM_n$ denote the n-th lower polariton (LP) branch and cavity mode (CM, black dashed lines), respectively. (c) The absorption (black line) and photoluminescence (red line) spectra of the TTPSB microbelt. (d) Scheme of the mode splitting in the region of the anticrossing point of the H-polarization mode (blue lines) and V-polarization mode (red lines) with opposite parity. Two opposite circularly polarized modes ($\Psi_{\sigma+}$ and $\Psi_{\sigma-}$) are obtained due to the anticrossing.

Two sets of modes with distinctive curvatures, marked respectively by black and orange lines, are observed (Fig. 1b). We further performed polarization-dependent angle-resolved reflectivity of the same sample by adding a linear polarizer in the detection optical path. It turns out that these two sets of modes are orthogonally linearly polarized. For example, the modes with larger curvatures are horizontal (H)-polarized (perpendicular to X-direction), while the modes with smaller curvatures are vertical (V)-polarized (parallel to X-direction, see Fig. 1a and Supplementary Fig. S2). This is due to the strongly polarized nature of excitons in TTPSB microbelts, which is supported by the anisotropy of the excitonic absorption in these microbelts (Supplementary Fig. S3). The excitons at the first excited singlet state ($S_1$ at 587 nm) of TTPSB, as shown in the absorption spectrum (blue line in Fig. 1c), can strongly couple only with the cavity photon modes that are V-polarized, which are in very good agreement with the lower polariton (LP) dispersions calculated by using the coupled harmonic oscillator (CHO) model (see Materials and Methods in Supplementary).[36] In sharp contrast, the excitonic resonance at 587 nm does not affect at all the H-polarized modes, which can be perfectly fitted by the planar cavity modes (Fig. 1b and 1c).

**Anticrossing and PL spectra.** More profoundly, the $LP_3$ (red line) and $CM_4$ (black dash line) modes as shown in Fig.1b anticross, forming tilted Dirac cones in the vicinity of the angle θ = ± 50°, because of the recently discovered emergent optical activity (OA),[37] arising when two orthogonally polarized modes with opposite parity are tuned near resonance. It provides an effective Zeeman field, changing sign with wave vector. It lifts the original crossing modes (Fig.1b). Theoretically, the OA-induced anticrossing can be efficiently described by an effective 2×2 Hamiltonian.[35] However, the strong mediation of the exciton mode that results in the normal mode splitting (e.g., polaritons) can optimize the system's Hamiltonian, i.e., a more accurate 4×4 Hamiltonian with the strong coupling of excitons and photons can be used to describe such a system. When the exciton mode strongly couples to only one of the orthogonally linearly polarized photon modes (V-mode), the effective 4×4 Hamiltonian in the linear polarization basis in this scenario has the form

$$H = \begin{pmatrix} E_{0,V} & 0 & \Omega/2 & 0 \\ 0 & E_{0,H} & 0 & 0 \\ \Omega/2 & 0 & E_{P,V} + \frac{\hbar^2 k^2}{2m} + \beta_0(k_x^2 - k_y^2) & -2i\beta_0 k_x k_y + \xi k_x \\ 0 & 0 & 2i\beta_0 k_x k_y + \xi k_x & E_{P,H} + \frac{\hbar^2 k^2}{2m} - \beta_0(k_x^2 - k_y^2) \end{pmatrix} \quad (1)$$

Here, $E_{0,H}$ ($E_{0,V}$) is the exciton energy in H- (V-) polarization, $E_{P,H}$ ($E_{P,V}$) is the ground energy of the H- (V-) polarized photon mode, $\Omega$ denotes the coupling of the cavity photon mode and the exciton mode, which occurs only in V-polarization, $m$ is the effective mass of cavity photons, $\boldsymbol{k}$ is the wavevector, $\beta_0$ represents the TE-TM splitting, and $\xi$ describes the emergent OA. The two lower modes calculated from the Hamiltonian (1) agree very well with the experimental results, as shown in Fig. 2a. The two anticrossing regions correspond to tilted Dirac cones, a particular case of topologically nontrivial valleys exhibiting opposite Berry curvature. They are characterized by opposite valley Chern numbers.

As analyzed above, the linear polarizations of the two anticrossing modes are exchanged at higher wave vectors. Take the lower branch as an example: the linearly V-polarized mode at larger angle transforms into the linearly H-polarized mode at smaller angle across the right-handed circularly polarized mode ($\Psi_{\sigma-}$) (c.f. Fig. 1d). Therefore, the linearly polarized photoluminescence (PL) can be emitted from the center of the reciprocal space, whereas oppositely circularly polarized emission should be detected in the anticrossing regions. In order to verify the validity of the anticipation, the angle-resolved measurement of the PL spectra of the same sample has been carried out and shown in Fig. 2a. Clearly, the PL emission distributes uniformly over the entire detected angle following the simulated polariton dispersions as indicated by the lines, evidencing that the angle-resolved PL signals originate from the polariton emission. Here, the positions of two LP emission spots match well with the spontaneous 0-0 and 0-1 PL transitions of the TTPSB microbelt, respectively (red line in Fig. 1c). We further experimentally measured the Stokes vector of the signals (see details in Supplementary). The $S_1$ components of the Stokes vector of the LP$_2$ and LP$_3$ branches show radically different polarization distributions (Fig. 2b). The LP$_2$ branch exhibits strong linearly polarized PL emission in the entire branch; however, the linear polarization of the LP$_3$ branch changes sign across the anticrossing regions. In comparison, the $S_3$ component of the Stokes vector shows that there are very weak circularly polarized signals in the LP$_2$ branch, whereas the circularly polarized signals in the LP$_3$ branch reach the maximal values at the anticrossing points and change sign at $k_x = 0$ because of the time-reversal symmetry (Fig. 2c). We also measured the two-dimensional (2D) Stokes vector of the LP$_3$ branch in the reciprocal space (Fig. 2d and 2e). It is clearly seen that in the $S_1$ component the value is zero and changes sign around the anticrossing regions, while in the same regions in the $S_3$ component the value reaches the maximum, but with opposite signs. We note that this kind of property occurs only along the $k_x$ direction determined by the orientation of the optical axis of the material. These results are in very good agreement with the theoretical results in Fig. 2f and 2g.

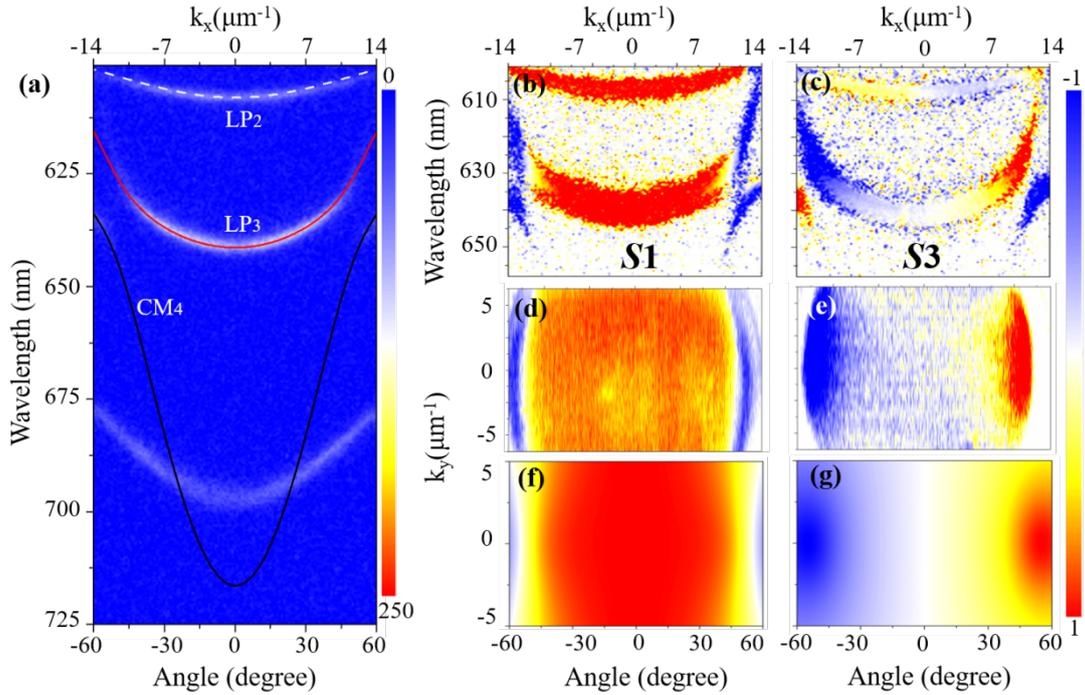

**Figure 2. Angle-resolved PL spectrum and Stokes parameters.** (a) Angle-resolved PL spectrum measured upon excitation of a 400-nm femtosecond laser at a low fluence. The LP$_3$ mode is marked by the red line and the CM$_4$ mode is marked by the black line. Measured Stokes parameters $S_1$ (b) and $S_3$ (c) of LP$_2$ and LP$_3$ modes. 2D maps of the Stokes parameters $S_1$ (d) and $S_3$ (e) of the LP$_3$ mode extracted from Fig. 2a. (f,g) Theoretically calculated Stokes parameters corresponding to (d,e), respectively. The two solid lines in (a) and the numerical results in (f,g) are obtained by solving the Hamiltonian (1) with the parameters: $E_{0,V} = E_{0,H} = 2.112$ eV, $E_{P,V} = 2.412$ eV, $E_{P,H} = 1.73$ eV, $\Omega = 590$ meV, $m = 0.78 \times 10^{-4} m_e$ ($m_e$ is the free electron mass), $\beta_0 = 0.5$ meV $\mu m^2$, and $\xi = 5.2$ meV $\mu m$.

**Polariton lasing through vibration-assisted relaxation.** It is well-known that photophysical properties of the organic microcrystals are strongly influenced by their aggregation effects. Benefiting from their H-aggregates (see the corresponding structural and spectroscopic analysis in Supplementary Fig.S4 and S5), TTPSB microbelts exhibit good lasing gain behavior in general, because their large Stokes shift minimizes the self-absorption, and the dipole-allowed 0-1 emission transition facilitates the lasing emission process (Supplementary Fig.S6). Importantly, the polaritons in the TTPSB crystal-filled microcavity possess large exciton fraction (according to the fitting parameters of the $LP_n$ (n = 1, 2, 3, and 4) branches at θ = 0° which corresponds to zero in-plane wavevector (Supplementary Table S2)) due to the positive detunings between the cavity modes and the exciton mode, which facilitates the bosonic condensation.[34]

In order to study the behavior of polariton lasing, the fluence of the pumping 400-nm femtosecond laser is increased to above the threshold ($P_{th}$) and the corresponding angle-resolved PL spectrum is shown in Fig. 3a. With the increase of the pump fluence to 1.5 $P_{th}$, the polariton intensity of $LP_3$ exhibits a sharp increase by about one order of magnitude near the two anticrossing points compared with that below the threshold, whereas the $LP_2$ emission increases only a little (see Supplementary Fig. S7). The blue line in Fig. 3b shows the integrated PL intensity of $LP_3$ emission as a function of the pump fluence, showing a typical threshold curve. The intensity dependence is separately fitted to power laws $x^p$ with $p$ = 0.35 ± 0.01, 3.55 ± 0.04, and 0.40 ± 0.02, respectively. The threshold of the polariton lasing is $P_{th}$ = 59.6 μJ/cm$^2$ at the first intersection between the sublinear and superlinear regions. Meanwhile, the full width at half-maximum (FWHM) of the $LP_3$ emission (red line) dramatically narrows from 2.75 nm below the threshold to 1.75 nm above the threshold. In sharp contrast, the integrated PL intensity of the $LP_2$ emission presents linear and slow increase (green line), indicating that the polaritons in the $LP_2$ branch remain in the spontaneous emission regime with unchanged FWHM. Fig. 3c shows the time-resolved PL from the $LP_3$ branch measured by a streak camera. At a very low pump density of 0.2 $P_{th}$, the polariton PL follows a single-exponential decay with a lifetime of τ = 0.36 ± 0.01 ns.

Upon increasing the pump density, a biexponential decay is observed, with the short component ascribed to bimolecular quenching.[38] Above the threshold, the PL decay time collapses to less than 30 ps, indicating the bosonic stimulation of the relaxation process,[39] and is limited by the resolution of our apparatus. All the above features confirm the polariton lasing at the $LP_3$ branch. The exact coincidence of the $LP_3$ branch with the 0-1 transition of TTPSB microbelts further suggests that the vibration-assisted relaxation mechanism selectively populates the $LP_3$ branch from the exciton reservoir through emitting a vibron.[34,40] Due to the lower quality factor of our cavity (Q ~ 300), the polariton condensation does not occur at the ground state of the $LP_3$ branch, but rather in the "bottleneck" region.

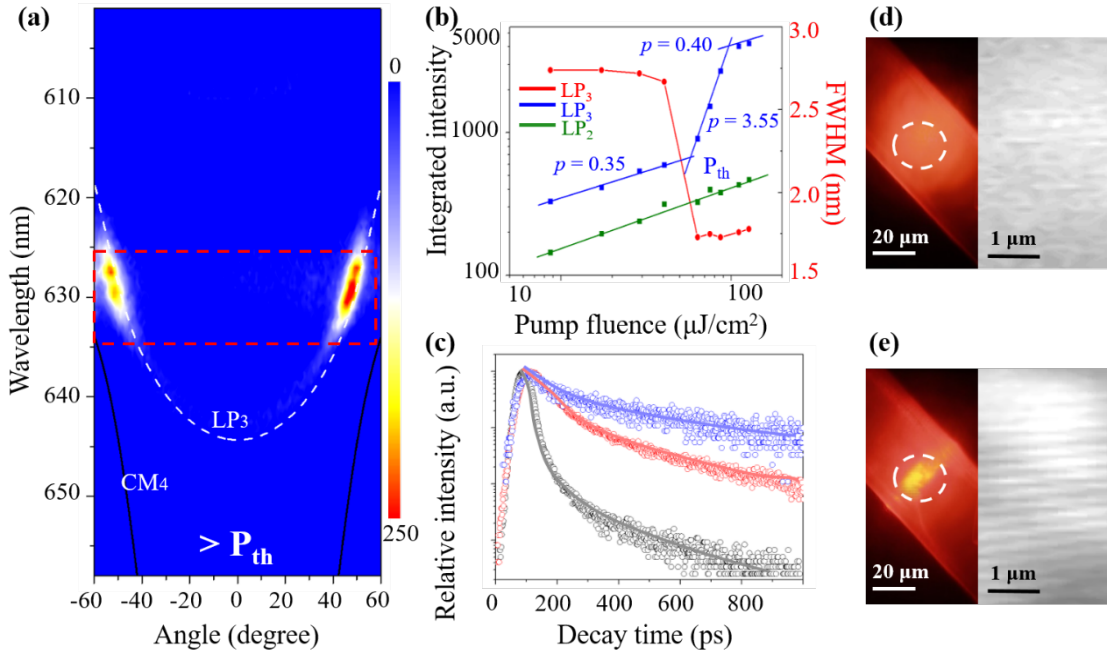

**Figure 3. Polariton lasing emission.** (a) Angle-resolved PL spectrum above the threshold. (b) Emission intensities of the $LP_2$ (green) and $LP_3$ (blue) branches and the FWHM (red) of the $LP_3$ branch as a function of the pump fluence. (c) PL decay profiles over time at different pump fluence: 0.2 $P_{th}$ (blue), 0.6 $P_{th}$ (red), and 1.2 $P_{th}$ (black). Real-space images recorded at 0.6 $P_{th}$ (d) and 1.2 $P_{th}$ (e) and the corresponding spatial coherence images recorded by a Michelson interferometer.

The long-range spatial coherence is also an important defining feature for polariton lasing. To demonstrate it, we sent the real-space PL image into a Michelson interferometer, in which one arm is a retroreflector to invert the PL image in a centro-symmetrical configuration. When the PL emission image and its inverted one are superimposed, no clear interference fringes can be observed at the pump fluence of 0.6 $P_{th}$ (Fig. 3d), representing the spontaneous regime. However, when the pump fluence exceeds the threshold, clear interference fringes can be well resolved throughout the whole detection region (Fig. 3e), confirming that the long-range spatial coherence is established in this case.

**Helical polariton lasing from topological valleys.** To understand the polarization properties of the polariton laser, we further performed the circularly polarized (σ+ and σ−) angle-resolved PL spectroscopy. Expectedly, the polariton lasing emission exhibits strong left-handed circular polarization in the vicinity of one anticrossing point of about -50° (Fig.4a), whereas a right-handed circularly polarized polariton laser can been obtained in the vicinity of the other anticrossing point of about +50° (Fig.4b). We further measured the Stokes vector, and it shows that in the $S_3$ component the maximal polarization degree is observed in the vicinity of two anticrossing points (Fig. 4d). The oppositely circularly polarized lasers are topologically protected by pseudo-spin time-reversal symmetry originating from the integrated SOC effects in the organic crystalline microcavity, giving rise to topological valleys. It is worth noting that the luminescence dissymmetry factor ($g_{lum}$) of our polariton laser is as high as 1.90, according to the equation, $g_{lum} = 2 \times (I_{\sigma+} - I_{\sigma-})/(I_{\sigma+} + I_{\sigma-})$, where $I_{\sigma+}$ and $I_{\sigma-}$ correspond to the laser output of left- and right-handed polarization, respectively.[41] This highly dissymmetric $g_{lum}$ stemming from topological valleys is far exceeding the reported circularly polarized light from organic systems, which may offer a new strategy for amplifying chiroptical response of organic small emissive molecules. Fig. 3e shows angular focusing of the circularly polarized emission β (defined as the ratio between the intensity emitted from the valleys $I_{\sigma+}$ or $I_{\sigma-}$ and the total intensity) of the laser output as a function of the pump fluence. Interestingly, β values exhibit a significant increase from 27% to 41% for σ+ and from 26% to 39% for σ-. This prominent chirality amplification benefits from the rapid population of polaritons into the anticrossing regions in the polariton lasing process. Fig. 4f shows the emission in 2D reciprocal space above the lasing threshold. It is clear that the lasing emission mainly focusses in the vicinity of the anticrossing points, which demonstrates the occupation of the polaritons in the two topological valleys.

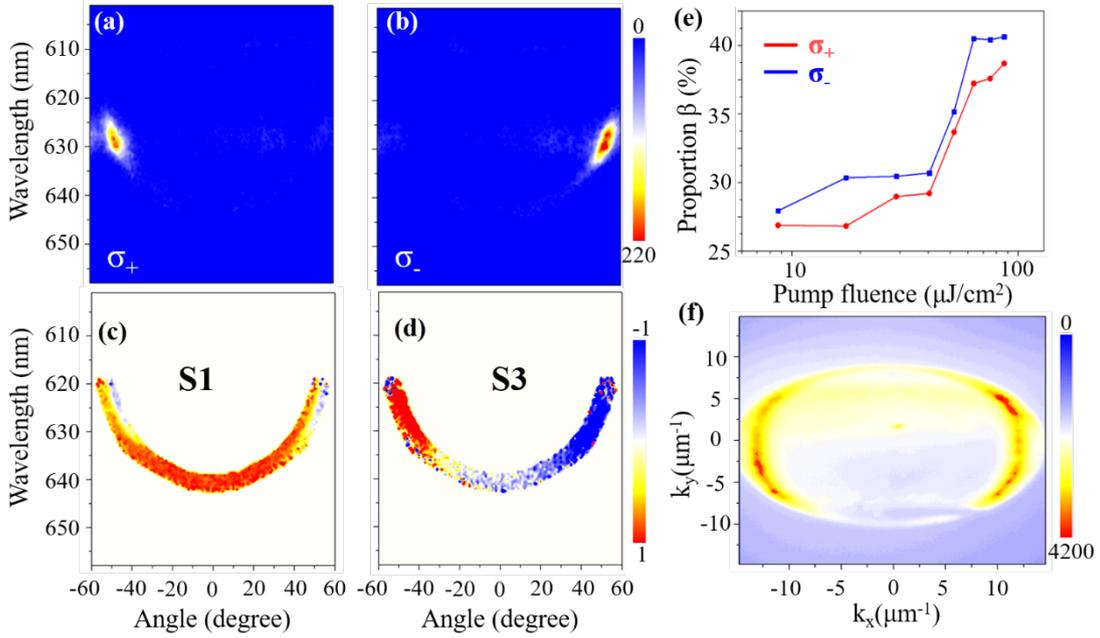

**Figure 4. Circular polarization distribution ($\sigma_+$ and $\sigma_-$).** (a) Angle-resolved PL spectra with left-handed (a) and right-handed (b) circular polarization above the threshold. Measured Stokes parameters $S_1$ (c) and $S_3$ (d) of the polariton laser from the $LP_3$ branch, corresponding to (a) and (b), respectively. (e) Circularly polarized proportion β of the output laser as a function of the pump fluence. (f) Emission in 2D reciprocal space above the lasing threshold, demonstrating the occupation of the polaritons in the two topological valleys.

**Conclusions**

In conclusion, we demonstrate helical topological polariton lasers in an organic anisotropic microcrystalline cavity by strongly coupling Frenkel excitons with the local nontrivial band geometry. Experimental and theoretical results indicate that the oppositely circularly polarized polariton lasers emit along different angular directions. The strong interaction of the excitonic component comes into play in bosonic condensation and the photonic component offers topological valleys. The left-handed and right-handed circularly polarized lasing is topologically protected by pseudo-spin time-reversal symmetry, originating from the optical activity of crystal structures. The significantly enhanced helical characteristics of output light are achieved by the nonlinear condensation process. Helical topological polariton lasers might provide a platform for the exploration of topological phenomena involving light-matter interaction and the development of polariton-based spintronic devices.

**Reference**

1   Hasan, M. Z. & Kane, C. L. Colloquium: Topological insulators. *Rev. Mod. Phys.* **82**, 3045-3067 (2010).

2   Weng, H., Yu, R., Hu, X., Dai, X. & Fang, Z. Quantum anomalous Hall effect and related topological electronic states. *Adv. Phys.* **64**, 227-282 (2015).

3   Klitzing, K. v., Dorda, G. & Pepper, M. New Method for High-Accuracy Determination of the Fine-Structure Constant Based on Quantized Hall Resistance. *Phys. Rev. Lett.* **45**, 494-497 (1980).

4   König, M. *et al.* Quantum Spin Hall Insulator State in HgTe Quantum Wells. *Science* **318**, 766-770 (2007).

5   Jotzu, G. *et al.* Experimental realization of the topological Haldane model with ultracold fermions. *Nature* **515**, 237-240 (2014).

6   Süsstrunk, R. & Huber, S. D. Observation of phononic helical edge states in a mechanical topological insulator. *Science* **349**, 47-50 (2015).

7   Wang, Z., Chong, Y., Joannopoulos, J. D. & Soljacic, M. Observation of unidirectional backscattering-immune topological electromagnetic states. *Nature* **461**, 772-775 (2009).

8   Hafezi, M., Mittal, S., Fan, J., Migdall, A. & Taylor, J. M. Imaging topological edge states in silicon photonics. *Nat. Photon.* **7**, 1001-1005 (2013).

9   Rechtsman, M. C. *et al.* Photonic Floquet topological insulators. *Nature* **496**, 196-200 (2013).

10  Haldane, F. D. & Raghu, S. Possible realization of directional optical waveguides in photonic crystals with broken time-reversal symmetry. *Phys. Rev. Lett.* **100**, 013904 (2008).

11  Zhao, H. *et al.* Non-Hermitian topological light steering. *Science* **365**, 1163-1166 (2019).

12    St-Jean, P. *et al.* Lasing in topological edge states of a one-dimensional lattice. *Nat. Photon.* **11**, 651-656 (2017).

13    Barik, S. *et al.* A topological quantum optics interface. *Science* **359**, 666-668 (2018).

14    Mittal, S., Goldschmidt, E. A. & Hafezi, M. A topological source of quantum light. *Nature* **561**, 502-506 (2018).

15    Blanco-Redondo, A. *et al.* Topological protection of biphoton states. *Science* **362**, 568-571 (2018).

16    Bahari, B. *et al.* Nonreciprocal lasing in topological cavities of arbitrary geometries. *Science* **358**, 636-640 (2017).

17    Bandres, M. A. *et al.* Topological insulator laser: Experiments. *Science* **359**, eaar4005 (2018).

18    Klembt, S. *et al.* Exciton-polariton topological insulator. *Nature* **562**, 552-556 (2018).

19    Karzig, T., Bardyn, C.-E., Lindner, N. H. & Refael, G. Topological Polaritons. *Phys. Rev. X* **5**, 031001(2015).

20    Nalitov, A. V., Solnyshkov, D. D. & Malpuech, G. Polariton Z topological insulator. *Phys. Rev. Lett.* **114**, 116401 (2015).

21    Liu, W. *et al.* Generation of helical topological exciton-polaritons. *Science* **370**, 600-604 (2020).

22    Weisbuch, C., Nishioka, M., Ishikawa, A. & Arakawa, Y. Observation of the coupled exciton-photon mode splitting in a semiconductor quantum microcavity. *Phys. Rev. Lett.* **69**, 3314-3317 (1992).

23    Kasprzak, J. *et al.* Bose-Einstein condensation of exciton polaritons. *Nature* **443**, 409-414 (2006).

24    Carusotto, I. & Ciuti, C. Quantum fluids of light. *Rev. Mod. Phys.* **85**, 299-366 (2013).

25    Jung, M., Fan, Z. & Shvets, G. Midinfrared Plasmonic Valleytronics in Metagate-Tuned Graphene. *Phys. Rev. Lett.* **121**, 086807 (2018).

26    Janot, A., Rosenow, B. & Refael, G. Topological polaritons in a quantum spin Hall

cavity. *Phys. Rev. B* **93** (2016).

27  Rodrigues, S. P. *et al.* Intensity-dependent modulation of optically active signals in a chiral metamaterial. *Nat. Commun.* **8** (2017).

28  Lodahl, P. *et al.* Chiral quantum optics. *Nature* **541**, 473-480 (2017).

29  Ayuso, D. *et al.* Synthetic chiral light for efficient control of chiral light–matter interaction. *Nat. Photon.* **13**, 866-871 (2019).

30  Cohen, Y. T. A. E. Enhanced Enantioselectivity in Excitation of Chiral Molecules by Superchiral Light. *Science* **332**, 333-336 (2011).

31  Sofikitis, D. *et al.* Evanescent-wave and ambient chiral sensing by signal-reversing cavity ringdown polarimetry. *Nature* **514**, 76-79 (2014).

32  Yuan, Z. *et al.* Predicting Gas Separation through Graphene Nanopore Ensembles with Realistic Pore Size Distributions. *ACS Nano* **15**, 1727-1740 (2021).

33  Sun, C. *et al.* Three-Dimensional Cuprous Lead Bromide Framework with Highly Efficient and Stable Blue Photoluminescence Emission. *Angew. Chem. Int. Ed.* **59**, 16465-16469 (2020).

34  Ren, J. *et al.* Efficient Bosonic Condensation of Exciton Polaritons in an H-Aggregate Organic Single-Crystal Microcavity. *Nano Lett.* **20**, 7550-7557 (2020).

35  Ren, J. H. *et al.* Nontrivial band geometry in an optically active system. *Nat. Commun.* **12**, 689 (2021).

36  Kena Cohen, S., Davanco, M. & Forrest, S. R. Strong exciton-photon coupling in an organic single crystal microcavity. *Phys. Rev. Lett.* **101**, 116401 (2008).

37  Rechcinska, K. *et al.* Engineering spin-orbit synthetic Hamiltonians in liquid-crystal optical cavities. *Science* **366**, 727-730 (2019).

38  Baldo, M. A., Holmes, R. J. & Forrest, S. R. Prospects for electrically pumped organic lasers. *Phys. Rev. B* **66** (2002).

39  Roumpos, G., Nitsche, W. H., Hofling, S., Forchel, A. & Yamamoto, Y. Gain-induced trapping of microcavity exciton polariton condensates. *Phys. Rev. Lett.* **104**, 126403 (2010).

40  Zasedatelev, A. V. *et al.* A room-temperature organic polariton transistor. *Nat. Photon.* **13**, 378-383 (2019).

41  Sun, C. L. *et al.* Lasing from Organic Micro-Helix. *Angew. Chem. Int. Ed.* **59**, 11080-11086 (2020).

**Acknowledgements**

This work was supported by the National Key R&D Program of China (Grant No. 2018YFA0704805, 2018YFA0704802 and 2017YFA0204503), the National Natural Science Foundation of China (22090022, 21833005, 21873065, 21790364 and 21673144), the Natural Science Foundation of Beijing, China (KZ202110028043), Beijing Talents Project (2019A23), the Open Fund of the State Key Laboratory of Integrated Optoelectronics (IOSKL2019KF01), Capacity Building for Sci-Tech Innovation-Fundamental Scientific Research Funds, Beijing Advanced Innovation Center for Imaging Theory and Technology. The Paderborn group acknowledges support by the Deutsche Forschungsgemeinschaft (DFG) through the collaborative research center TRR142 (project A04, No. 231447078).

The authors thank Dr. H.W. Yin from ideaoptics Inc. for the support on the angle-resolved spectroscopy measurements.

**Author contributions**

T.L., J.-H.R., and Q.L. designed the experiments and performed experimental measurements. X.M., S.S., G.M. and D.S. performed the theoretical calculation and analysis. T.L., X.M., F.L., Q.L., S.S., G.M., D.S. and H.-B.F. wrote the manuscript with contributions from all authors. Q.L. and H.-B.F. supervised the project. All authors analyzed the data and discussed the results.

**Competing interests**

The authors declare no competing interests.

**Additional information**

Correspondence should be addressed to Q.L.: liaoqing@cnu.edu.cn

Supporting Information

# Helical polariton laser from nontrivial band geometry in organic crystalline microcavity

Teng Long,[1] Xuekai Ma,[2] Jiahuan Ren,[1,3] Feng Li,[4] Qing Liao,[1,*] Stefan Schumacher,[2,5] Guillaume Malpuech,[6] Dmitry Solnyshkov,[6,7] Hongbing Fu[1]

[1]Beijing Key Laboratory for Optical Materials and Photonic Devices, Department of Chemistry, Capital Normal University, Beijing 100048, People's Republic of China
E-mail: liaoqing@cnu.edu.cn

[2]Department of Physics and Center for Optoelectronics and Photonics Paderborn (CeOPP), Universität Paderborn, Warburger Strasse 100, 33098 Paderborn, Germany

[3]Tianjin Key Laboratory of Molecular Optoelectronic Science, School of Chemical Engineering and Technology, Collaborative Innovation Center of Chemical Science and Engineering (Tianjin), Tianjin University, Tianjin 300072, P. R. China

[4]Key Laboratory for Physical Electronics and Devices of the Ministry of Education & Shaanxi Key Lab of Information Photonic Technique, School of Electronic and Information Engineering, Xi'an Jiaotong University, 710049 Xi'an, China.

[5]Wyant College of Optical Sciences, University of Arizona, Tucson, Arizona 85721, United States

[6]Institut Pascal, PHOTON-N2, Université Clermont Auvergne, CNRS, SIGMA Clermont, F-63000 Clermont-Ferrand, France

[7]Institut Universitaire de France (IUF), 75231 Paris, France

## MATERIALS AND METHODS

### 1. Synthesis of TTPSB

The compound used in our work, 1,4-dimethoxy-2,5-di(2,2',5',2''-ter-thiophenestyryl) benzene (TTPSB) was synthesized according to Horner-Wadsworth-Emmons reaction (F. Gao *et al.*, *Angew. Chem. Int. Ed.* 2010, 49, 732-735. & Z. Xu *et al.*, Adv. Mater. 2012, 24, OP216-OP220.). All starting materials were purchased from Alfa Aesar and used as received without further purification. The tetrahydrofuran (THF, HPLC grade) and hexane were purchased from Beijing Chemical Agent Ltd., China. Ultra-pure water with a resistance of 18.2 MΩ·cm$^{-1}$ were used in all experiments, produced by Milli-Q apparatus (Millipore).

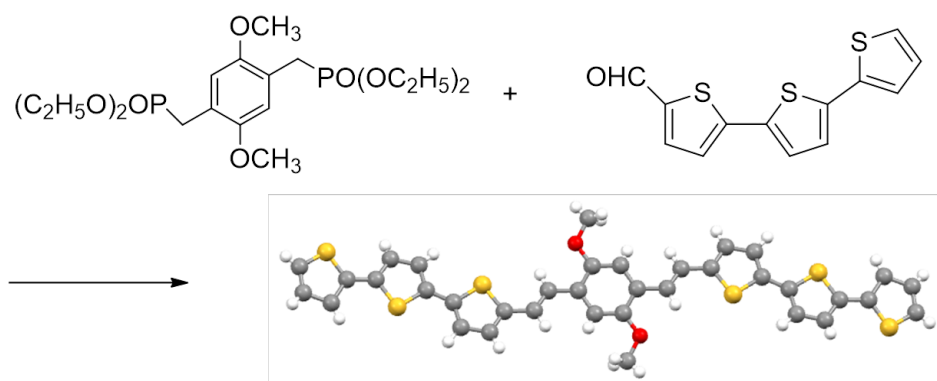

**Scheme S1.** The synthesis of the TTPSB molecule.

Without water and oxygen, a mixture of 2,5-bismethoxy-1,4-xylene-bis(diethyl phosphonate) (510 mg, 1.16 mmol) and catalyzer NaH (54.72mg, 2.28mmol) in tetrahydrofuran (THF) solution was cooled in an ice bath at 0 °C during a 30 min period. Then the 2, 2'-Bithiophene-5-carboxaldehyde (620.0 mg, 2.26 mmol)/THF solution inject reaction bulb. (Scheme S1) And the reaction mixture was stirred at room temperature for 6 hours and subsequently poured into a little water, spin dry. Then, pouring the orange powders into the Buchner funnel and wash it with alcohol. Finally orange powder was obtained as the compound (425.2 mg, 0.62 mmol) in 53% yield. 1H NMR (400 MHz, C4D8O): δ 7.41-7.32 (m, 4 H), 7.28-7.22 (m, 4H), 7.2 (s, 2H), 7.18-7.13 (m, 6 H), 7.04-7.01 (m, 4H), 3.91 (s, 6 H); MS (MALDI-TOF): 682.0.

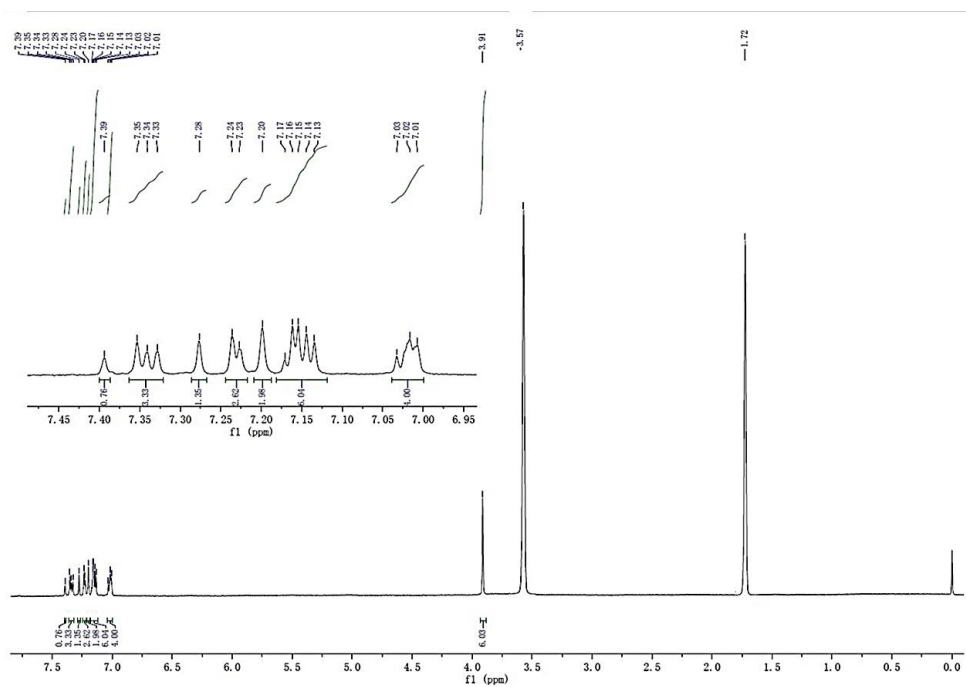

$^1$H Nuclear magnetic resonance （NMR) spectrum of TTPSB.

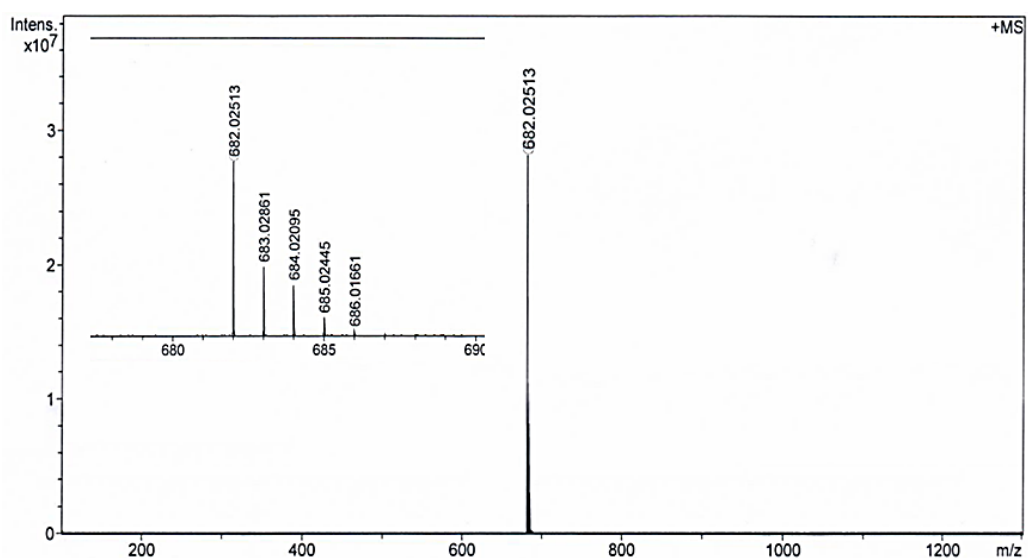

MALDI-MS spectrum of TTPSB.

## 2. The preparation of TTPSB microbelts

In our experiment, TTPSB microbelts were fabricated using a facile physical vapor deposition (PVD) method. A quartz boat carrying 3 mg TTPSB was then placed in the center of a quartz tube which was inserted into a horizontal tube furnace. A continuous flow of cooling water inside the cover caps was used to achieve a temperature gradient over the entire length of the tube. To prevent oxidation of TTPSB, Ar was used as inert gas during the PVD process (flowrate: 15 sccm·min$^{-1}$). The pre-prepared hydrophobic substrates were placed on the downstream side of the argon flow for product collection and the furnace was heated to the sublimation temperature of TTPSB (at temperature region of ~ 320 °C), upon which it was physically deposited onto the pre-prepared hydrophobic substrates at temperature region of ~ 230 °C for 1 hours.

## 3. The preparation of TTPSB microcavity

Firstly, we use the metal vacuum deposition system (Amostrom Engineering 03493) to thermally evaporate silver film with the thickness of 85 ± 5 nm (reflectivity R ≥ 99%) on the glass substrate, the root mean square roughness (Rq) of the silver film in the 5 μm × 5 μm area is 2.45 nm, a 20 ± 2 nm $SiO_2$ layer was deposited using vacuum electron beam evaporate on the silver film with $R_q$ of 2.31 nm, the deposited rates were both 0.2 Å/s and the base vacuum pressure is 3×10$^{-6}$ Torr. This silver/$SiO_2$ film composite structure was placed as a substrate in a horizontal tube furnace for sample deposition. The TTPSB microbelts were uniformly dispersed on the silver/$SiO_2$ film substrate. Then 20 ± 2 nm $SiO_2$ and 35 ± 2 nm (R ≈ 50%) silver was fabricated to form the microcavity. The 20-nm $SiO_2$ layers is used to prevent the fluorescence quenching of TTPSB microbelts caused by directly contact of the metallic silver with the crystal.

**4. Structural and spectroscopic characterization**

As-prepared TTPSB microbelts were characterized by field emission scanning electron microscopy (FE-SEM, HITACHI S-4800) by dropping on a silicon wafer. Samples examined by transmission electron microscopy (TEM, JEOL, JEM-2100) were obtained by one drop of the solution being dropped on a carbon-coated copper grid and evaporated. TEM measurement was performed at room temperature at an accelerating voltage of 100 kV. The X-ray diffraction (XRD, Japan Rigaku D/max-2500 rotation anode X-ray diffractometer, graphite monochromatized Cu K$_\alpha$ radiation ($\lambda$ =1.5418 Å)) operated in the 2θ range from 3 to 30°, by using the samples on a cleaned glass slide.

The fluorescence micrograph, diffused reflection absorption and emission spectra were measured on Olympus IX71, HITACHI U-3900H, and HITACHI F-4600 spectrophotometers, respectively. Fluorescence quantum yield ($\Phi$) of TTPSB monomer solution in THF measured through a relative method by using Rhodamine 6G as a standard and $\Phi$ of TTPSB nanowires measured through an absolute method by using an integration sphere. The photoluminescence spectra of isolated single TTPSB microbelt in microcavity was characterized by using a homemade optical microscope equipped with a 50 × 0.9 NA objective (Scheme S2). The second harmonic ($\lambda$ = 400 nm, pulse width 150 femtosecond) of a 1 kHz Ti: sapphire regenerative amplifier was focused to a 50-μm diameter spot to excite the selected single TTPSB on a two-dimensional (2D) movable table. Spatially resolved PL spectra were collected underneath by using a 3D-movable objective and detected using a liquid-nitrogen cooled charge-coupled device (CCD).

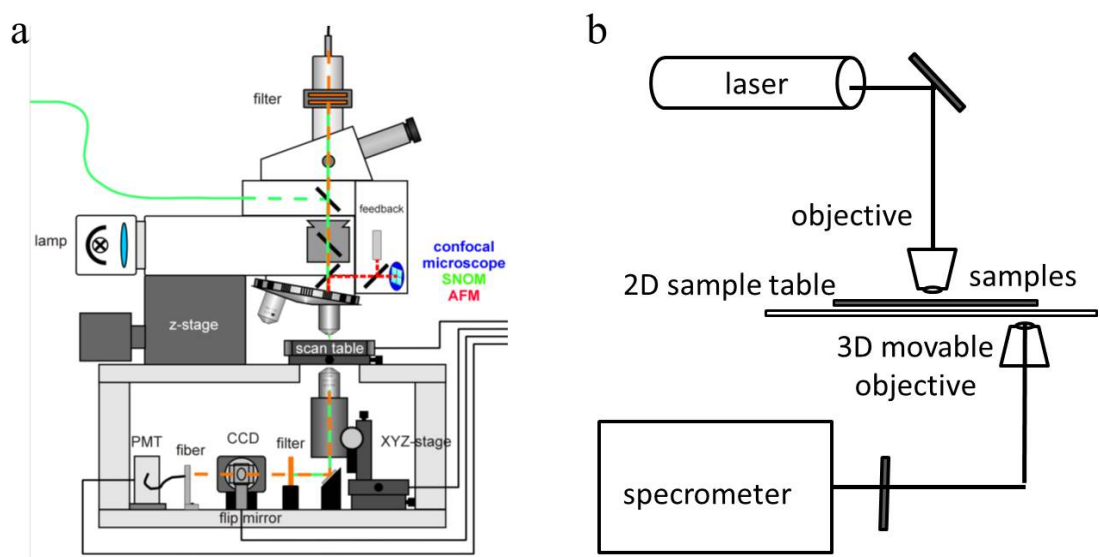

**Scheme S2.** Schematic demonstration of the experimental setup for the optical characterization: (a) the near-field scanning optical microscopy, and (b) the transmittance optical path for the waveguide measurements.

### 5. The angle-resolved spectroscopy characterization

The angle-resolved spectroscopy was performed at room temperature by the Fourier imaging using a 100× objective lens of a NA 0.95, corresponding to a range of collection angle of ±60° (Scheme S3). An incident white light from a Halogen lamp with the wavelength range of 400-700 nm was focused on the area of the microcavity containing a TTPSB microbelt. The k-space or angular distribution of the reflected light was located at the back focal plane of the objective lens. Lenses L1-L4 formed a confocal imaging system together with the objective lens, by which the k-space light distribution was first imaged at the right focal plane of L2 through the lens group of L1 and L2, and then further imaged, through the lens group of L3 and L4, at the right focal plane of L4 on the entrance slit of a spectrometer equipped with a liquid-nitrogen-cooled CCD. The use of four lenses here provided flexibility for adjusting the magnification of the final image and efficient light collection. Tomography by scanning the image (laterally shifting L4) across the slit enabled obtaining spectrally resolved two-dimensional (2D) k-space images.

In order to investigate the polarization properties, we placed a linear polarizer, a half-wave plate and a quarter-wave plate in front of spectrometer to obtain the polarization state of each pixel of the k-space images in the horizontal-vertical (0° and 90°), diagonal (±45°) and circular (σ+ and σ− ) basis (S. Dufferwiel *et al.*, *Phys. Rev. Lett.* 2015, 115, 246401. & F. Manni *et al.*, *Nat. Commun.* 2013, 4, 2590.).

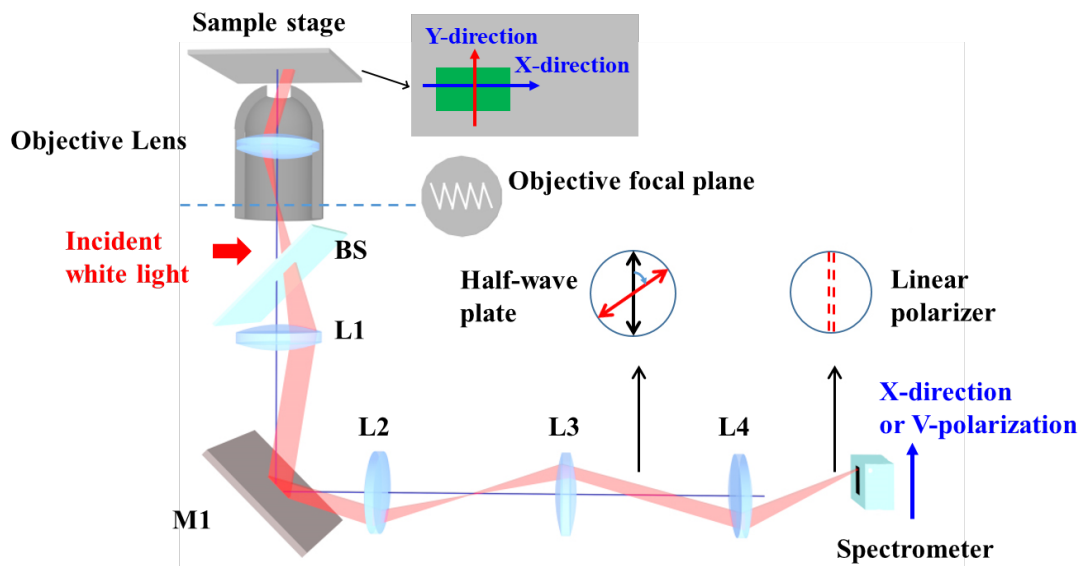

**Scheme S3.** Experimental setup allowing to obtain polarization-resolved complete state tomography. BS: beam splitter; L1-L4: lenses; M1: mirror. The red beam traces the optical path of the reflected light from the sample at a given angle.

## 6. Polariton dispersion

The polariton dispersion in Fig. 1b was calculated by a coupled harmonic oscillator Hamiltonian (CHO) model (S. Kena-Cohen *et al.*, *Phys. Rev. Lett.* 2008, 101, 116401.). The 2×2 matrix in equation (1) below describes the CHO Hamiltonian:

$$\begin{pmatrix} E_{CMn}(\theta) & \Omega/2 \\ \Omega/2 & E_X \end{pmatrix} \begin{pmatrix} \alpha \\ \beta \end{pmatrix} = E \begin{pmatrix} \alpha \\ \beta \end{pmatrix} \quad (1)$$

Where $\theta$ represents the polar angle, $E_{CMn}(\theta)$ is the cavity photon energy of the $n^{th}$ cavity mode as a function of $\theta$, $E_X$ is the exciton 0–0 absorption energy of TTPSB microbelts at 2.11 eV (587 nm) and $\Omega$ (eV) denotes the coupling. The magnitudes $|\alpha|^2$ and $|\beta|^2$ correspond to the photonic and the excitonic fraction, respectively.

The cavity photon dispersion is given by:

$$E_{CMn}(\theta) = \sqrt{\left(E_c^2 \times \left(1 - \frac{\sin^2\theta}{n_{eff}^2}\right)^{-1}\right) - (n-1) \times l} \quad (2)$$

where $E_c$ represents the cavity modes energy at $\theta = 0°$, $E_{CM1}(\theta)$ represents the energy of the first cavity mode when n=1, $(n-1) \times l$ represents the energy difference from the first cavity mode. The effective refractive index ($n_{eff}$ = 1.8 and 3) is extracted from the fitting results. The theoretical fitting dispersion of the uncoupled cavity modes ($n_{eff}$ = 1.8, black solid line) and coupled cavity modes ($n_{eff}$ = 3, black dash line) is shown in Figure 1b. Diagonalization of this Hamiltonian yields the eigenvalues, $E_{\pm}(\theta)$, which represents the upper and lower polariton (UP and LP) in-plane dispersions (H. Deng *et al.*, *Rev. Mod. Phys.* 2010, 82, 1489-1537.),

$$E_{\pm}(\theta) = \frac{E_X + E_{CMn}(\theta)}{2} \pm \frac{1}{2}\sqrt{\left(E_X - E_{CMn}(\theta)\right)^2 + \hbar^2\Omega^2} \quad (3)$$

**Figure S1**

Scanning and transmission electron microscopy (SEM and TEM) depict that typical TTPSB microbelts with smooth surface and sharp edge (inset in Figure S1a) have been successfully fabricated on a large scale (Figure S1a). Their length (*l*) is found to range between 40 and 60 μm, while the width (*w*) and the height (*h*) is determined to be about 40 μm and 720 nm, respectively, according to atomic force microscopy (AFM) measurements (Figure S1c and d). The observed sharp spots in selected area electron diffraction (SAED) pattern (Figure S1b) clarifies that these microbelts are single crystalline in nature. According to that monoclinic TTPSB crystals belong to the space group of P2$_1$/n, with cell parameters of a = 10.4006 Å, b = 5.5454 Å, c = 27.0574 Å, α = γ = 90°, and β = 93.409° (CCDC No. 1554859, Table S1), the blue triangle spot in Figure S1b is ascribed to (103) Bragg reflections with a *d*-spacing value of 13.558 Å, and the red square spot corresponds to (011) crystal plane with *d*-spacing value of 11.328 Å, in good agreement with the cell parameters of this monoclinic crystal structure.

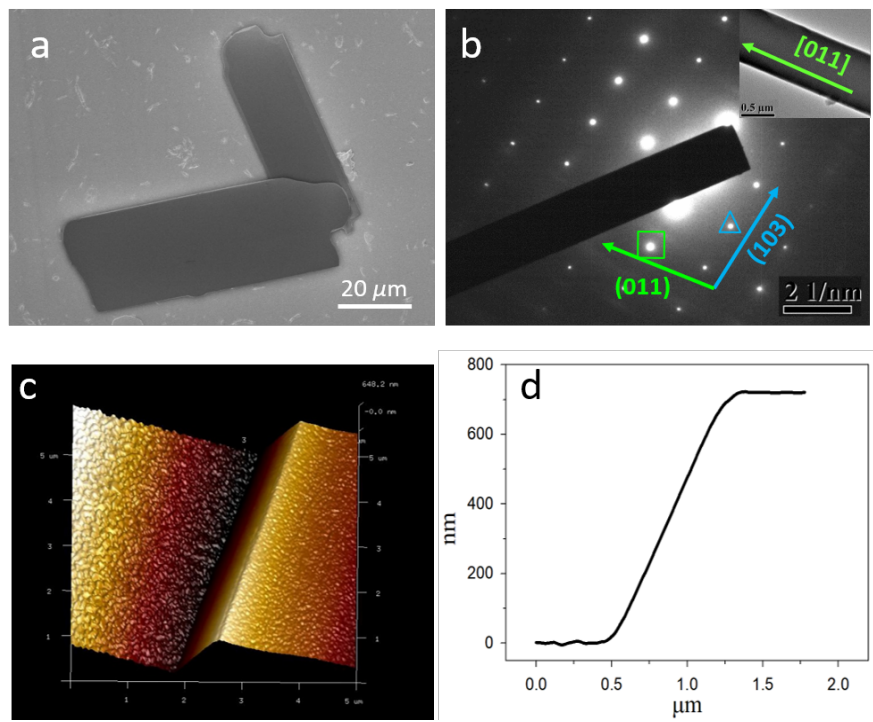

**Figure S1.** (a) SEM image of as-prepared TTPSB microbelts. Inset: the end of a single microbelt in the high magnification. (b) SAED pattern recorded by directing the electron beam perpendicular to the top-surface of a single TTPSB microbelt. Inset: the corresponding TEM image. (c) AFM image of as-prepared TTPSB microbelts. (d) The graph shows the topography line profile measured across the microbelt.

**Table S1.** Crystal date and structure refinement for TTPSB.

| Space Group | P $2_1/n$ |
|---|---|
| $a$ | 10.4006Å |
| $b$ | 5.54540Å |
| $c$ | 27.0574Å |
| $\alpha$ | 90.000° |
| $\beta$ | 93.409° |
| $\gamma$ | 90.000° |

**Figure S2**

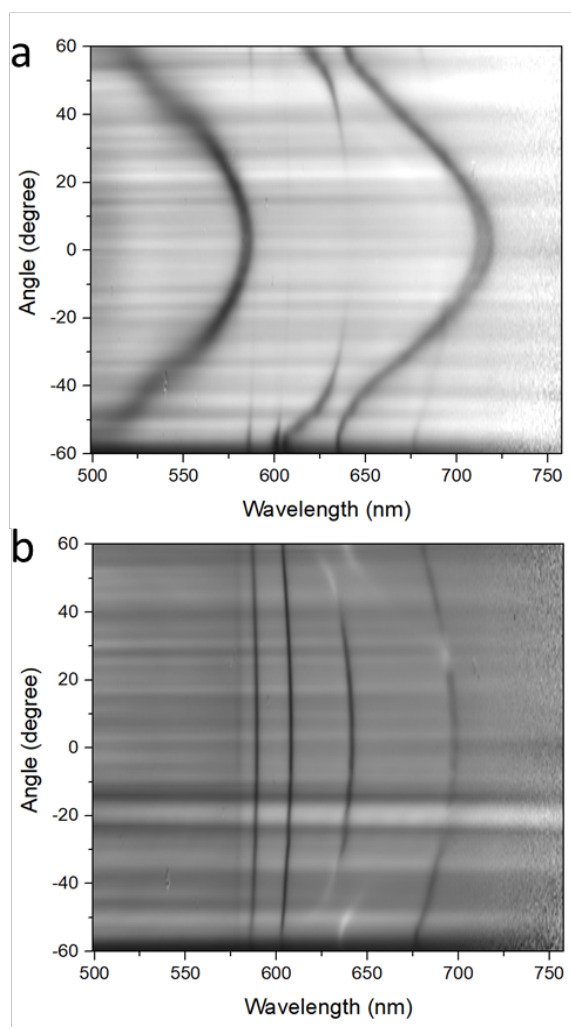

**Figure S2.** Measured k-space angle-resolved reflectivity spectra of a selected microcavity at room temperature in horizontal (H) polarization (a) and vertical (V) polarization (b) along X-direction of the single-crystal cavity.

**Figure S3**

A broad and intense absorption peak is observed (blue line) when the polarization of the white light from a halogen lamp is adjusted to be parallel to the X-direction of the microbelt, whereas the absorption is much weaker (red line) when the polarization of the white light is perpendicular to the X-direction of the microbelt. The descriptions of polarization directions also see Supplementary Scheme S3. These distinct polarization-dependent absorption features are consistent with the fact that the anisotropy is a result of the highly ordered uniaxial alignment of TTPSB molecules in single crystalline microbelts.

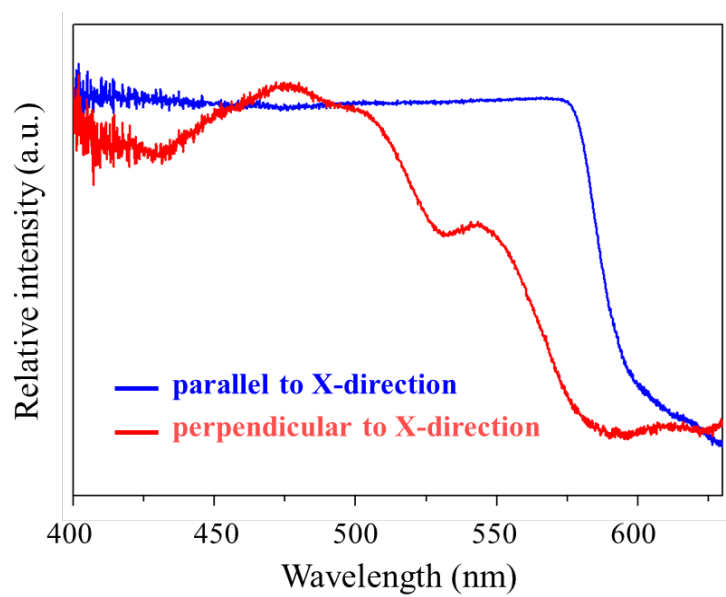

**Figure S3.** Polarization-dependent absorption spectra of a single TTPSB microbelt.

**Figure S4**

It can be seen from Figure S4a that X-ray diffraction (XRD) spectra (red line) of ensemble microbelts deposited on a silicon substrate is dominated by the (002) and (10-3) series of peaks compared to XRD spectrum of TTPSB powder (black line), suggesting that these crystal facets are the main exposed surfaces of microbelts. Taking into account TEM image (inset of Figure S1b), it can be concluded that TTPSB microbelts grow preferentially along the crystal [011] direction. We also simulated the growth shape of TTPSB crystal based on the attachment energies using Material Studio package (D. Winn *et. al*, *AIChE J.* 2000, 46, 1348-1367.). We found that the predicted thermodynamic stable morphology is also belt-like structure, and the preferential growth along the crystal [011] direction is also the main stacking direction predicted (Figure S4b), which agrees with the results of the observed 1D microbelts (Figure S1a). According to Kasha's exciton model (M. Kasha, *Discuss. Faraday Soc.* 1950, 9, 14-19.), intermolecular herringbone packing generally advocates H-aggregation.

Combining the analysis of SAED, TEM and XRD results, TTPSB molecules within 1D microbelts adopt a herringbone packing arrangement and stack co-facially along the crystal [011] plane with the shortest separation about 3.47 Å, which indicates a typical π-π stacking (Figure S4c). Further analyze the molecules arrangement (Figure S4d-g), the pitch angle (which defines the angle between the molecular transition dipole and the π-stack direction) of 82.2° (corresponding to the longitudinal displacements between neighboring molecules of 0.47 Å) ensures co-facially π-stacking, which might be beneficial to efficient charge transport channel for 1D-MWs. While the roll angle of 39.3° (corresponding to the transverse displacements 4.23 Å) greatly reduces the quenching effect caused by the π-π interaction and brings strong PL emission. According to the molecular exciton model (F. Würthner *et. al*, *Angew. Chem. Int. Ed.* 2011, 50, 3376-3410.), the pitch angle is <54.7° in a J-type aggregate and >54.7° in an H-type aggregate. Therefore, it is expected that H-type coupling occurs in TTPSB microbelts.

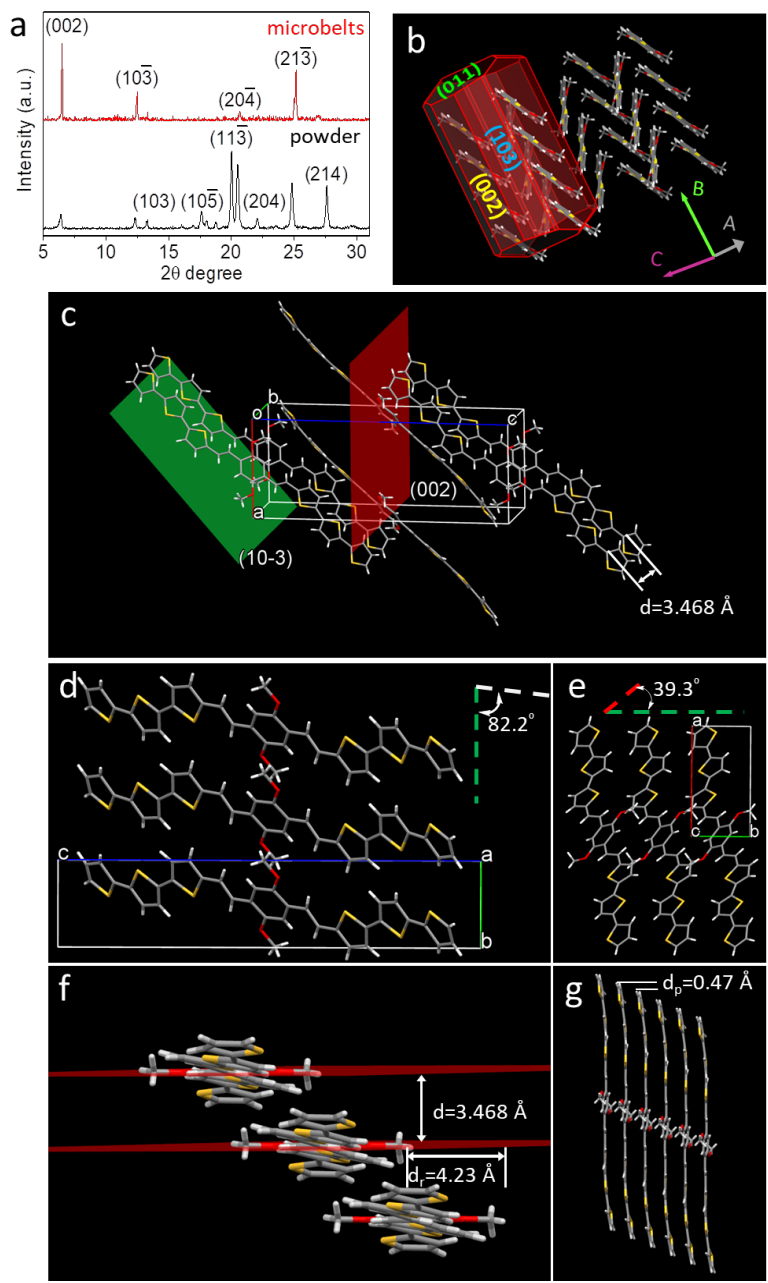

**Figure S4.** (a) XRD profiles of microbelts (red line) and powder (black line). (b) Simulated growth morphology of TTPSB molecules using Material Studio package. (c) Molecular packing arrangement of TTPSB in the nanowires, viewed almost along the crystal *b*-axis. (d) Viewed normal to *bc* plane with illustration of pitch angle. (e) Viewed normal to *ab* plane with illustration of roll angle. (f) Viewed the shortest separation of $d$ = 3.468 Å and the transverse displacement of $d_r$ = 4.23 Å. (g) Viewed the longitudinal displacement of $d_p$ = 0.47 Å.

**Figure S5**

Figure S5 depicts the absorption and photoluminescence (PL) spectra of TTPSB monomers in diluted tetrahydrofuran (THF) solution and ensemble microbelts placed on a quartz substrate. The related photophysical parameters were summarized in Table S2. The absorption spectrum of the diluted solution exhibits a broad featureless peak at 473 nm attributed to the lowest $S_0 \rightarrow S_1$ transition of the monomers. PL emission shows clearly vibrationally structure at 538, 567, and 611 nm with contributions arising mainly from vinyl stretching mode (top panel of Figure S5a). In intense contrast, the maximum absorption spectrum of microbelts shows a slight blue-shift to 470 nm, with additional bands around 543 nm (2A) and 566 nm (1A) due to aggregate states in microbelts (bottom panel of Figure S5a). Their PL spectrum is dominated by 0-1 transition and maximum emission posits at 621 nm, due to the self-absorption caused by the overlap between their absorption and their emission. This is a specific fingerprint of H-type aggregates with a "face-to-face" molecular-packing arrangement.

TTPSB monomers has poorly emission with the PL quantum yield (Φ) of $0.04 \pm 0.01$ through a relative method by using Rhodamine 6G as a standard, while microbelts exhibit a moderate value Φ of $0.15 \pm 0.01$ through an absolute method by using an integration sphere (Table S2). This is a typical "aggregation-induced emission (AIE)". To further investigate the nature of the excited states, we performed time-resolved fluorescence measurements by single photon counter (Figure S5b and S5c). The monomer solution emission at 540 nm decays monoexponentially, yielding a lifetime of $\tau_m = 182 \pm 1.4$ ps. The PL decay of microbelts at 625 nm was also fitted monoexponentially with a time constant of $\tau_{microbelts}$ = 425 ± 2.3 ps, which is much longer than that of TTPSB monomers. Based on the equation of $k_r = \Phi/\tau$ (M. Gsänger *et. al*, *Angew. Chem. Int. Ed.* 2010, 49, 740-743.), the radiative decay rates ($k_r$) are calculated to be $k_{r,m}$ = 0.220 ns$^{-1}$ and $k_{r,microbelts}$ = 0.353 ns$^{-1}$ for monomers and microbelts, respectively (Table S2). The restriction of rotation motion of terthiophenestyryl substituents in solid state were considered to enhance the nonradiative decay and result in AIE effect.

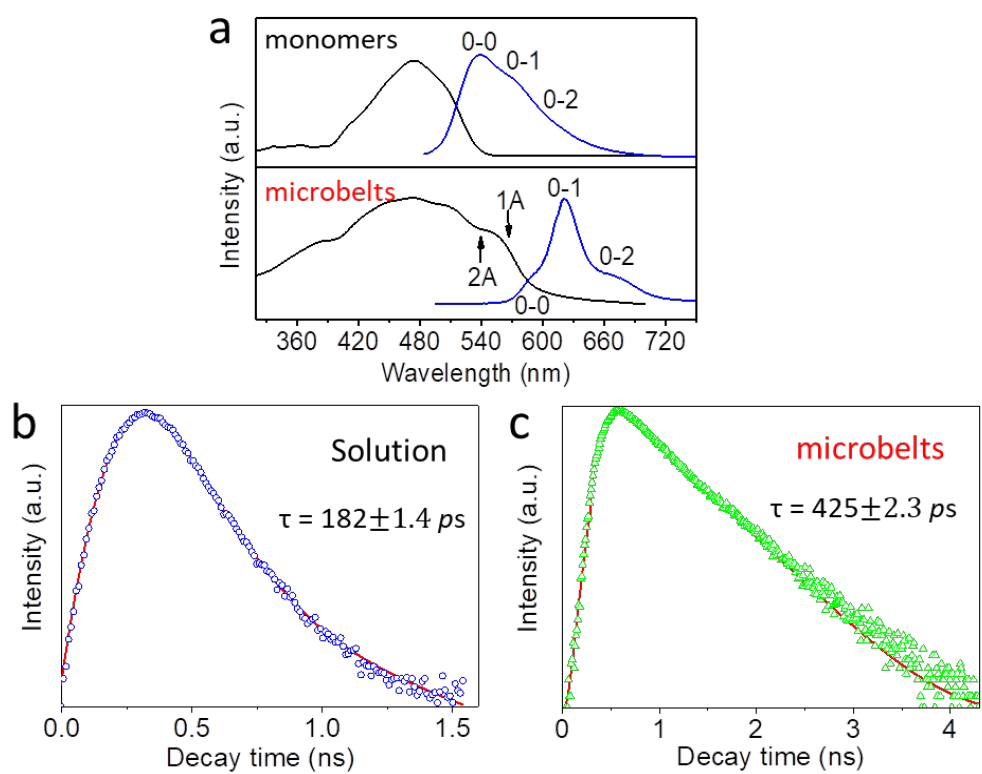

**Figure S5.** (a) Normalized absorption (black line) and PL (blue line) spectra of ensemble microbelts placed on a glass substrate (lower) and monomers in THF solution (upper). PL decay profiles of TTPSB monomers in THF solution (b) and ensemble microbelts on Silica substrate (c).

**Table S2.** Photophysical parameters of TTPSB monomers in the dilute solution and microbelts.

| Sample | $\lambda_{abs}$ (nm) | $\lambda_{em}$ (nm) | $\Phi^a$ | $\tau^b$ (ps) | $k_r^c$ (ns$^{-1}$) |
|---|---|---|---|---|---|
| Monomer | 473 | 538<br>567<br>611 | 0.04 | 182±1.4 | 0.220 |
| microbelts | 470<br>543<br>566 | 583<br>631<br>666 | 0.15 | 425±2.3 | 0.353 |

$^a\Phi$ of monomer solution in THF measured through a relative method by using Rhodamine 6G as a standard and $\Phi$ of microbelts measured through an absolute method by using an integration sphere. $^b$Fluorescence lifetime. $^c$Radiative decay rate calculated according to $k_r = \Phi/\tau$.

**Figure S6**

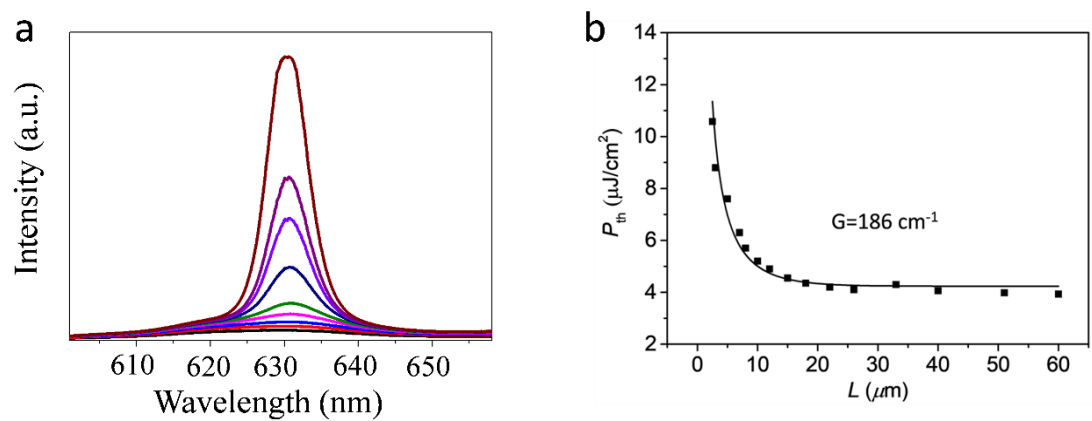

**Figure S6.** (a) Amplified spontaneous emission from individual TTPSB microbelts. The red box represents the gain region. (b) Threshold values as a function of TTPSB microbelt lengths, for example: 2.3 $\mu$m, 3 $\mu$m, 5 $\mu$m, 7 $\mu$m, 8 $\mu$m, 10 $\mu$m, 12 $\mu$m, 15 $\mu$m, 18 $\mu$m, 22 $\mu$m, 26 $\mu$m, 33 $\mu$m, 40 $\mu$m, 51 $\mu$m, and 60 $\mu$m.

**Table S3. Coupled Harmonic Oscillator Model Fitting Results for LP$_1$ to LP$_4$.**

| Coupling mode | LP1 | LP2 | LP3 | LP4 |
|---|---|---|---|---|
| Rabi splitting (meV) | 140 | 520 | 590 | 592 |
| Detuning (eV) | 1.10 | 0.65 | 0.30 | -0.1 |
| $|\alpha|^2$ | 0.01 | 0.09 | 0.27 | 0.58 |
| $|\beta|^2$ | 0.99 | 0.91 | 0.73 | 0.42 |

**Figure S7**

At low pump fluence of $P = 0.3\ P_{th}$, the LP dispersions of both $LP_2$ and $LP_3$ branches exhibit a broad and uniform emission distribution at all angles as shown in Figure 2a, which follows the simulated polariton dispersions, proving that the angle-resolved PL signal indeed originates from polariton emissions. The fitted full width at half-maximum (FWHM) is 2.90 nm (Fig. S7b).

As the pump fluence increases, the $LP_3$ polariton intensity at $\theta = -40°$ exhibit rapid increase near two anticrossing points, whereas the $LP_2$ emission remains a little growth. When the pump fluence reaches $P = 1.1\ P_{th}$, the obvious new peak at 633 nm with FWHM of about 1.75 nm, which shows a slight blueshift compared to spontaneous PL emission at 635 nm, has been observed as shown in Fig. S7a and S7c. Further increasing the pump fluence, the intensities of the PL emission present the nonlinear increase as shown in Fig. 3b. The obviously blue-shifted PL peak, spectral narrowing (Fig. S7b-d), and nonlinear increase of the spectral intensity confirm that the polariton lasing has occurred at two anticrossing points in the $LP_3$ branch.

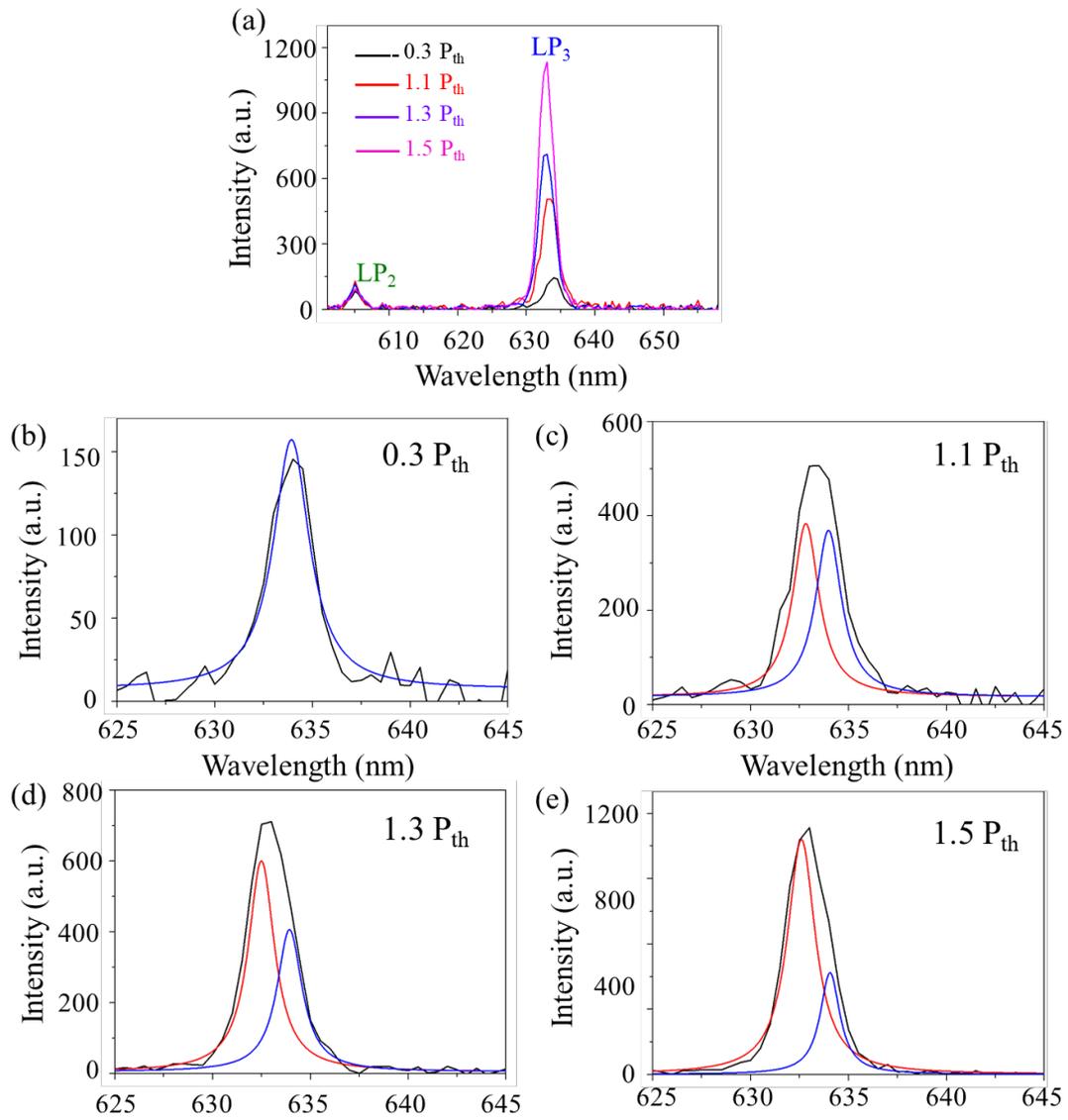

**Figure S7.** (a) PL intensities at θ = -50° as a function of pump fluence. (b-e) Fitted Lorentzian line shapes for the calculation of FWHM.